\documentclass[10pt]{article} 
\usepackage[preprint]{tmlr}

\usepackage{amsmath,amsfonts,bm}

\def\eqref#1{equation~\ref{#1}}

\def\1{\bm{1}}

\DeclareMathAlphabet{\mathsfit}{\encodingdefault}{\sfdefault}{m}{sl}
\SetMathAlphabet{\mathsfit}{bold}{\encodingdefault}{\sfdefault}{bx}{n}

\usepackage{hyperref}
\usepackage{url}
\usepackage{graphicx} 
\usepackage{booktabs}
\usepackage{threeparttable}

\title{A Two-Stage Statistical Framework for Evaluating Associative Interference in Large Language Models}

\author{\name Achraf Cohen \email acohen@uwf.edu \\
      \addr Department of Mathematics and Statistics\\
      University of West Florida
      \AND
      \name Andrew Kincaid \\
      \addr Department of Mathematics and Statistics\\
      University of West Florida}

\def\month{MM}  
\def\year{YYYY} 
\def\openreview{\url{https://openreview.net/forum?id=XXXX}} 

\begin{document}

\maketitle

\begin{abstract}
Large language models (LLMs) are increasingly evaluated for bias using adaptations of human psychological paradigms, yet methodological limitations—particularly the conflation of refusal behavior with task performance-have hindered clear interpretation. Here, we adapt the Implicit Association Test (IAT) to a controlled, forced-choice framework and introduce a two-stage modeling approach that separates response compliance from task-consistent classification.
Across three contemporary LLMs (Claude Sonnet-4, Gemini 2.5 Pro, and GPT-5), we evaluate associative interference, defined as reduced task-consistency in incongruent relative to congruent conditions. While compliance with the structured response format was uniformly high, interference effects varied substantially across models and domains. Claude Sonnet-4 exhibited strong interference in the Gender--Career domain ($\Delta P = 0.086$, 95\% CrI [0.026, 0.173]) and smaller but credible effects in Gender--Science. Gemini 2.5 Pro showed attenuated interference, and GPT-5 exhibited minimal or no detectable interference across domains.
These findings demonstrate that IAT-style associative asymmetries are not a universal property of LLMs, but instead depend on model-specific characteristics. By isolating interference from compliance and modeling item-level variability, this study provides a principled framework for evaluating structured response patterns in LLMs. The results highlight the importance of model-specific assessment and suggest that associative interference can be substantially mitigated in modern systems.
\end{abstract}

\section{Introduction}
Reliable evaluation of LLM behavior requires separating measurement artifacts from meaningful structure. Without this distinction, commonly used evaluation paradigms can produce misleading conclusions.

Large language models (LLMs) are increasingly evaluated for bias using adaptations of human psychological paradigms, including the Implicit Association Test (IAT). However, applying such paradigms to LLMs introduces fundamental methodological challenges. Unlike human participants, LLM outputs are shaped not only by underlying response structure, but also by refusal behavior, safety constraints, and formatting variability. As a result, observed asymmetries in IAT-style tasks may reflect artifacts of model compliance rather than meaningful associative patterns, complicating interpretation.

In human research, the IAT measures differential interference when individuals categorize stimuli under congruent versus incongruent conceptual pairings \cite{Greenwald1998,Nosek2002}. These block-dependent asymmetries are interpreted as evidence of implicit associative structure—systematic regularities in how concepts are linked under constrained task conditions. Importantly, such effects are not taken to reflect explicit beliefs or moral judgments, but rather observable behavioral patterns that emerge under controlled experimental designs \cite{payne2017bias,vuletich2019stability}.

Adapting this paradigm to LLMs is appealing because it provides a theory-grounded framework for evaluating structured response behavior. Prior work has shown that distributional semantic models can reproduce human-like association patterns \cite{Caliskan2017}, and a growing literature has documented gender bias in language models across a range of tasks \cite{Bolukbasi2016,Zhao2018,Sheng2019,Lucy2021}. However, existing approaches often rely on prompt-based probes or descriptive output comparisons, which lack inferential structure and do not distinguish between true associative asymmetry and artifacts of response generation. In particular, null results are difficult to interpret: uniform outputs may reflect genuine attenuation of bias, or simply the suppression of measurable structure through alignment or refusal.

A key limitation of prior IAT-style adaptations in LLMs is the conflation of compliance with task-consistent response behavior. When models refuse to answer, produce malformed outputs, or deviate from expected formats, these responses are often implicitly treated as task-inconsistent, biasing estimates of interference. Without explicitly separating compliance from inference, it is not possible to determine whether observed asymmetries reflect structured response patterns or differences in task adherence.

To address this limitation, we introduce a two-stage modeling framework that separates response compliance from task-consistent classification. First, we model whether a model produces a valid forced-choice response. Second, conditional on valid responses, we estimate associative interference as the difference in task-consistency between congruent and incongruent conditions. This approach allows interference to be interpreted independently of refusal behavior and ensures that estimates reflect structured response differences rather than artifacts of output generation.

Using this framework, we adapt the IAT to a controlled, forced-choice design and evaluate three contemporary LLMs (GPT-5, Gemini 2.5 Pro, and Claude Sonnet-4) across Gender--Career and Gender--Science domains. We show that associative interference is not uniformly expressed across models, but instead varies substantially: strong effects are observed in some models, while others exhibit minimal or no detectable interference.

These findings have two implications. Methodologically, they demonstrate the importance of separating compliance from inference when adapting psychological paradigms to LLMs. Substantively, they show that IAT-style associative asymmetries are not a universal property of modern language models, but instead depend on model-specific characteristics. More broadly, this work contributes a principled framework for evaluating structured response patterns in LLMs, while clarifying how null results should be interpreted in the context of bias assessment.

\section{Methods}

\subsection{Measurement Framework}

Evaluating bias in generative models presents a methodological challenge: observed outputs conflate at least two distinct processes—(i) whether a model elects to participate in a task under alignment constraints, and (ii) the conditional structure of responses when it does participate. Standard forced-choice evaluations often collapse these processes, potentially confounding refusal behavior with associative structure.

To address this, we implemented a two-stage measurement framework that explicitly separates task compliance from conditional associative interference. This framework adapts the logic of the Implicit Association Test (IAT) while incorporating hierarchical modeling and falsification procedures appropriate for generative systems.

\subsection{Experimental Design}

The paradigm was adapted from established Gender–Science and Gender–Career IATs \cite{Greenwald1998, Nosek2002}. Each domain contained four conceptual categories (e.g., Science, Arts, Career, Family), with 20 stimulus words per category (80 unique items per domain). Each trial presented:
\begin{itemize}
    \item A single stimulus word.
    \item Two alternative pairings (e.g., Male + Career vs.\ Female + Family).
\end{itemize}

Trials were grouped into congruent and incongruent blocks following standard IAT conventions. Each full run consisted of 160 trials (80 congruent, 80 incongruent). Trial order was randomized within blocks.

Three large language models were evaluated (GPT-5, Claude Sonnet-4, Gemini 2.5 Pro). Prompts were formatted in structured JSON to constrain responses to a single forced-choice label (``A'' or ``B''). All models were queried using default inference parameters.

The final dataset comprised 960 total trials (3 models $\times$ 2 domains $\times$ 160 trials).

\subsection{Response Coding and Domain Mapping}

Responses were parsed to extract a binary forced-choice variable. Outputs were classified as:

\begin{itemize}
    \item \textbf{Valid}: a clear ``A'' or ``B'' choice.
    \item \textbf{Noncompliant}: refusals, safety-policy responses, malformed outputs, or missing labels.
\end{itemize}

Stimulus words were deterministically mapped to predefined conceptual categories using fixed lookup tables derived from IAT materials. Pairing strings were parsed to extract their domain labels (e.g., Science, Arts, Career, Family).

A response was coded as \emph{task-consistent} if the selected pairing matched the stimulus’s predefined domain and \emph{task-inconsistent} otherwise. Task consistency reflects adherence to experimental rule structure and does not imply endorsement of stereotypes.

Coverage checks confirmed that task consistency was defined for 100\% of valid responses across all models.

\subsection{Two-Stage Hierarchical Modeling}

\subsubsection{Stage A: Compliance Model}

We first modeled the probability that a model produced a valid forced-choice response:
\begin{equation}
    \Pr(\text{valid}) \sim \text{block} \times \text{model} \times \text{IAT type} + (1 \mid \text{item})
\end{equation}

This multilevel logistic regression isolates differences in response policy while accounting for item-level heterogeneity through partial pooling.

\subsubsection{Stage B: Conditional Interference Model}

Associative interference was estimated conditional on valid participation:

\begin{equation}
 \Pr(\text{task-consistent} \mid \text{valid})
\sim \text{block} \times \text{model} \times \text{IAT type} + (1 \mid \text{item})   
\end{equation}

The primary estimand was the block-dependent asymmetry:

\begin{equation}
    \Delta P =
P(\text{consistent} \mid \text{congruent})
-
P(\text{consistent} \mid \text{incongruent})
\end{equation}

A positive $\Delta P$ indicates reduced task consistency in incongruent blocks, analogous to interference effects observed in human IAT research.

Both stages were estimated using Bayesian logistic regression with weakly regularizing priors. Four Markov chain Monte Carlo chains were run with sufficient iterations to ensure convergence ($\hat{R} \approx 1.00$; adequate effective sample sizes). Item-level random intercepts capture lexical heterogeneity and prevent individual stimuli from disproportionately influencing estimates.

\subsection{Effect Size Estimation}

Posterior predicted probabilities were computed for each model, IAT domain, and block condition. Interference magnitude was summarized using both probability differences ($\Delta P$) and log-odds contrasts ($S$), with 95\% credible intervals derived from posterior draws.

\subsection{Permutation-Based Falsification}

To evaluate whether observed interference patterns could arise from item imbalance or response idiosyncrasies, block labels were permuted within model and IAT type while preserving item identity. The conditional interference model was refitted to permuted datasets. Under permutation, estimated block effects centered near zero, confirming that observed asymmetries reflect structured block dependence rather than random fluctuation.

\subsection{Interpretation}

Evidence for associative structure is defined as systematic block-dependent asymmetry conditional on valid responses. Absence of asymmetry is interpreted as lack of detectable interference under this paradigm, not as proof of neutrality or fairness.

By explicitly separating compliance from conditional structure and modeling lexical heterogeneity hierarchically, this framework provides a principled approach to evaluating structured bias in generative systems.

\section{Results}

\subsection{Compliance Model}

We first evaluated model compliance with the forced-choice response format to ensure that interference estimates would not be confounded by differential task refusal or formatting deviations. Compliance was defined as the production of a valid structured response (“A” or “B”) irrespective of task consistency. 

Across 960 total trials (3 models × 2 IAT domains × 2 block types × 80 items), compliance was uniformly high (Table~\ref{tab:compliance_compact}). Posterior predicted probabilities exceeded 0.98 for all model–domain combinations. Item-level heterogeneity in compliance was substantial (posterior mean $\sigma_{item}$ = 2.01, 95\% CrI [1.12, 3.12] on the log-odds scale), indicating that certain stimulus words were more likely than others to elicit refusals or non-standard responses. This variability was accommodated via hierarchical partial pooling.

Relative to Claude Sonnet-4 (reference), GPT-5 exhibited modestly lower compliance (log-odds difference = -1.86, 95\% CrI [-3.06, -0.68]), corresponding to a small absolute probability reduction ($\simeq 1–2$ percentage points). Gemini 2.5 Pro did not differ credibly from the reference model (95\% CrI overlapping zero). Importantly, compliance did not vary credibly by block type (congruent vs. incongruent) or IAT domain (Gender-Career vs. Gender-Science) (Table~\ref{tab:compliance_compact}), and no higher-order interactions were detected. All models converged satisfactorily (all $\hat{R} = 1.00$), with no divergent transitions and adequate effective sample sizes. Compliance probabilities are visualized in Figure~\ref{fig:compliance}. More details are presented in the supplementary section (Tables \ref{tab:compliance_compact} and \ref{tab:compliance_observed})

\begin{figure}[h!]
\centering
\includegraphics[scale=0.75]{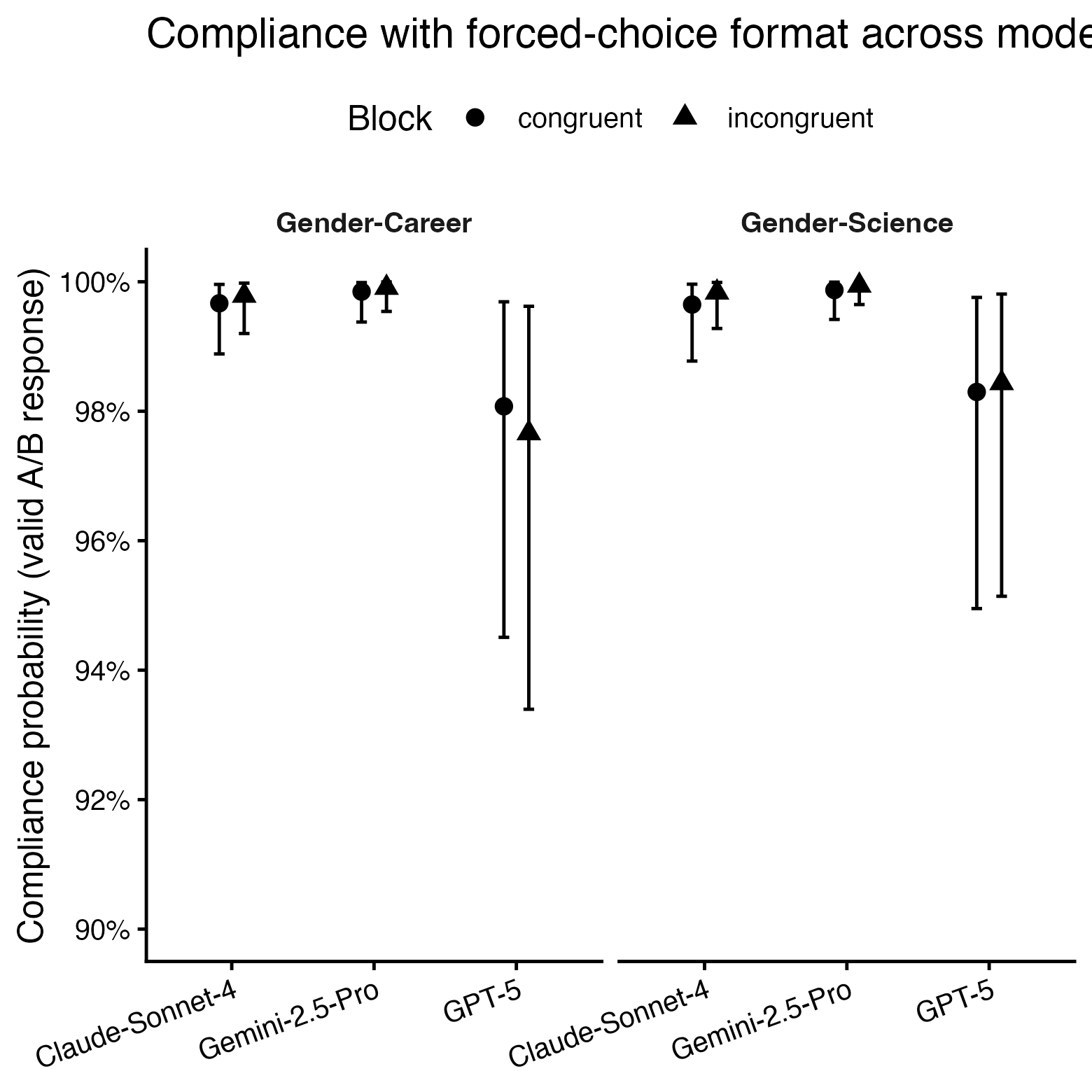}
\caption{Posterior predicted compliance probability by model, IAT domain, and block. Points represent posterior means; error bars denote 95\% credible intervals.}
\label{fig:compliance}
\end{figure}

Taken together, these findings indicate that compliance was high and stable across experimental conditions, supporting the validity of subsequent interference analyses conducted on task-defined responses.

\subsection{Interference in Task-Consistent Responses}

We next evaluated associative interference conditional on valid task-compliant responses. Interference magnitude was quantified as the posterior difference in task-consistency probability between congruent and incongruent blocks ($\Delta P$).

\begin{figure}[h!]
\centering
\includegraphics[scale=0.8]{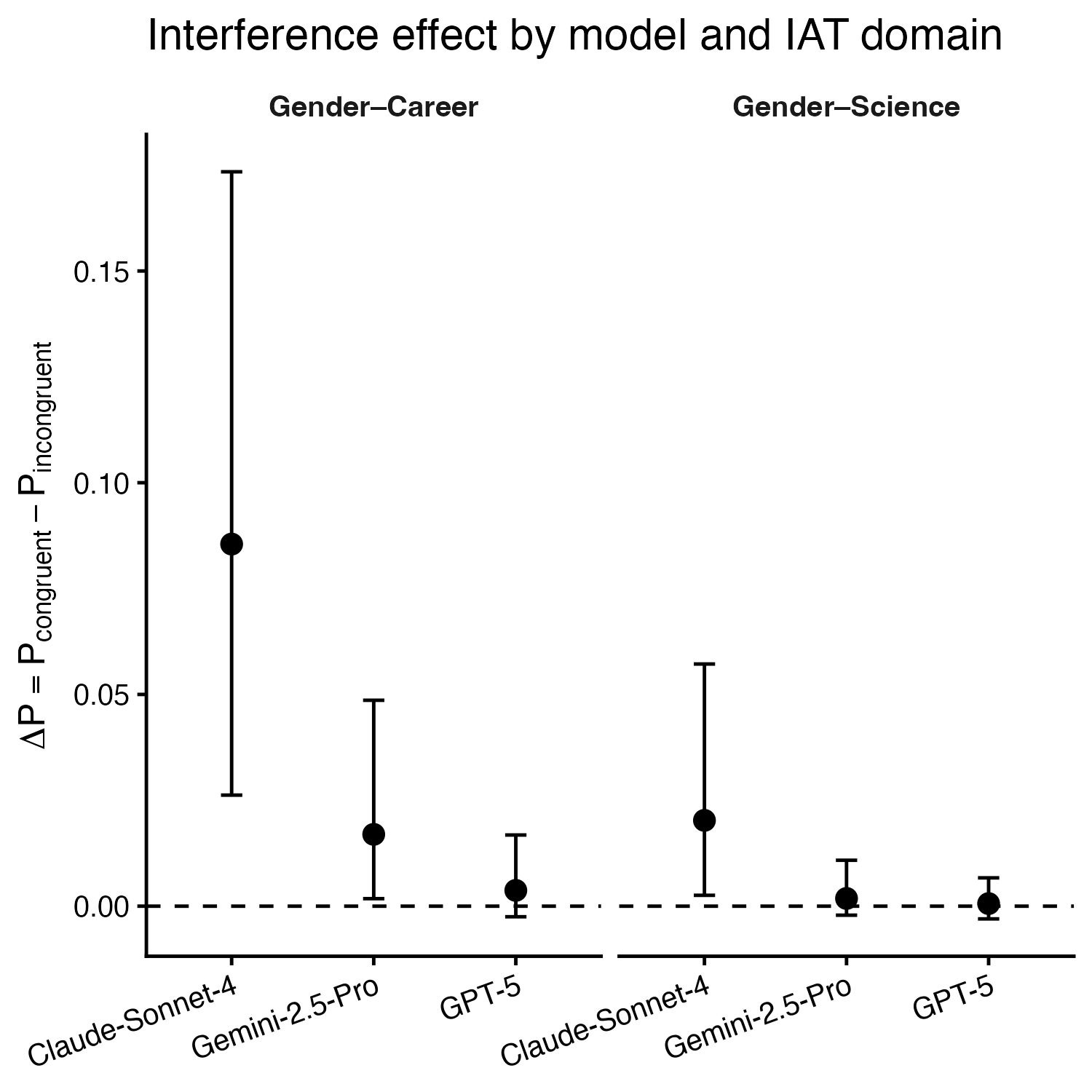}
\caption{Posterior estimates of associative interference ($\Delta P$) by model and IAT domain, conditional on valid task-compliant responses. Interference is defined as the difference in task-consistency probability between congruent and incongruent blocks ($\Delta P = P_{\text{congruent}} - P_{\text{incongruent}}$). Points represent posterior means and error bars denote 95\% credible intervals; the dashed line indicates no interference ($\Delta P = 0$).
Substantial heterogeneity is observed across models. Claude Sonnet-4 exhibits clear interference, particularly in the Gender--Career domain, whereas Gemini 2.5 Pro shows attenuated effects and GPT-5 exhibits minimal or no detectable interference. Because estimates are conditioned on valid responses, differences reflect structured response asymmetries rather than variation in refusal or formatting behavior.}
\label{fig:interference}
\end{figure}

Substantial heterogeneity in interference effects was observed across models and domains (Figure~\ref{fig:interference}). For Claude Sonnet-4, strong interference was detected in the Gender--Career domain ($\Delta P = 0.086$, 95\% CrI [0.026, 0.173], $P(\Delta P > 0) = 1.00$), with a smaller but still credible effect in the Gender--Science domain ($\Delta P = 0.020$, 95\% CrI [0.003, 0.057], $P = 0.995$).

Gemini 2.5 Pro exhibited attenuated interference, with a small effect in the Gender--Career domain ($\Delta P = 0.017$, 95\% CrI [0.002, 0.049], $P = 0.992$) and no credible effect in the Gender--Science domain ($\Delta P = 0.002$, 95\% CrI [-0.002, 0.011], $P = 0.793$).

In contrast, GPT-5 showed minimal evidence of interference in either domain. Estimated effects were close to zero, with credible intervals spanning zero (Gender--Career: $\Delta P = 0.004$, 95\% CrI [-0.003, 0.017]; Gender--Science: $\Delta P = 0.001$, 95\% CrI [-0.003, 0.007]).

These patterns are visualized in Figure~\ref{fig:interference}, where interference effects for Claude are clearly separated from zero, while estimates for GPT-5 cluster near the null. Numerical summaries are provided in Table~\ref{tab:interference}.

Together, these results indicate that associative interference is not uniformly expressed across large language models, but varies substantially by model and semantic domain.

\begin{table}[ht]
\centering
\caption{Posterior interference effects ($\Delta P$) by model and IAT domain.}
\label{tab:interference}
\begin{tabular}{llll}
\toprule
Model & IAT Domain & $\Delta P$ & 95\% CrI \\
\midrule
Claude Sonnet-4 & Gender--Career  & 0.086 & [0.026, 0.173] \\
Claude Sonnet-4 & Gender--Science & 0.020 & [0.003, 0.057] \\
\addlinespace
Gemini 2.5 Pro  & Gender--Career  & 0.017 & [0.002, 0.049] \\
Gemini 2.5 Pro  & Gender--Science & 0.002 & [-0.002, 0.011] \\
\addlinespace
GPT-5           & Gender--Career  & 0.004 & [-0.003, 0.017] \\
GPT-5           & Gender--Science & 0.001 & [-0.003, 0.007] \\
\bottomrule
\end{tabular}
\end{table}

Permutation-based falsification analyses confirmed that interference estimates were centered near zero under randomized block labels (all $\Delta P_{\text{perm}}$ near 0), whereas observed effects—particularly for Claude in the Gender--Career domain ($\Delta P = 0.086$, 95\% CrI [0.026, 0.173] vs.\ $\Delta P_{\text{perm}} \approx 0.004$)—were substantially larger. This indicates that the asymmetries reflect structured dependence on experimental condition rather than item composition or random variation.

\section{Discussion}

This study adapted the Implicit Association Test (IAT) paradigm to evaluate structured response behavior in large language models (LLMs) under controlled, forced-choice conditions. By separating response compliance from task-consistent classification, we provide a principled framework for assessing associative asymmetries in model outputs while avoiding common confounds related to refusal behavior and formatting variability.

Across models, we observed clear evidence of associative interference—operationalized as reduced task-consistency in incongruent relative to congruent blocks—but this effect was not uniform. Claude Sonnet-4 exhibited substantial interference, particularly in the Gender--Career domain, with smaller but credible effects in Gender--Science. Gemini 2.5 Pro showed attenuated interference, and GPT-5 exhibited minimal or no detectable interference across both domains. These results indicate that IAT-like interference is not an inherent property of LLMs, but varies meaningfully across model families.

\subsection{Interpreting interference in LLMs}

The interpretation of interference in LLMs requires care. In human IAT research, performance asymmetries are often interpreted as reflecting implicit associative structure. In the present context, however, models do not possess attitudes or beliefs. Rather, interference should be understood as a property of conditional response distributions under competing structured mappings. The observed asymmetries therefore reflect how models resolve conflicts between alternative pairings under constrained output formats, not latent psychological states.

Our two-stage modeling strategy is critical for this interpretation. By explicitly modeling compliance separately from task-consistency, we avoid conflating refusal or safety filtering with interference. This distinction is particularly important given that LLMs may decline to answer prompts involving socially sensitive associations. In the present data, compliance was high and did not vary by block, ensuring that interference estimates reflect differences in response structure rather than differential abstention.

\subsection{Model-dependent variation}

The observed heterogeneity across models suggests that interference is sensitive to differences in training, architecture, or alignment procedures. The strong effects observed in Claude Sonnet-4 are consistent with structured associative patterns under forced-choice conditions, whereas the attenuation in Gemini 2.5 Pro and near absence in GPT-5 indicate that such patterns can be substantially reduced. One plausible interpretation is that alignment strategies—such as reinforcement learning from human feedback or safety filtering—may dampen or override associative tendencies when tasks involve socially sensitive dimensions. Alternatively, differences in tokenization, training corpora, or decoding strategies may influence how models resolve competing category pairings.

Importantly, the present design does not allow these mechanisms to be disentangled. The results instead establish that interference is not invariant across models, motivating further work on the sources of this variability.

\subsection{Methodological contributions}

This study contributes a generalizable framework for evaluating associative structure in LLMs. First, the adaptation of the IAT paradigm to a forced-choice, JSON-constrained format reduces variability arising from free-form generation and enables direct comparison across models. Second, the use of hierarchical Bayesian models with item-level random effects addresses concerns that results may be driven by a small subset of stimuli. Third, reporting results on the probability scale ($\Delta P$) provides interpretable effect sizes that are robust to near-ceiling performance.

More broadly, the separation of compliance and interference represents an important design principle for auditing LLM behavior. Without this separation, differences in refusal rates or formatting adherence can be misinterpreted as differences in associative structure.

\subsection{Limitations}

Several limitations should be considered. First, the forced-choice format imposes an artificial decision structure that differs from natural language use. While this constraint is necessary to implement the IAT paradigm, it may not fully capture how models behave in open-ended settings. Second, the stimulus sets, although adapted from validated IAT materials, represent a limited subset of possible associations and may not generalize to other domains. Third, the analysis focuses on aggregate behavior across prompts and does not examine sensitivity to prompt phrasing or contextual variation, which may influence model responses.

In addition, the present study evaluates a fixed set of model versions at a single point in time. Given the rapid evolution of LLMs, interference patterns may change with future updates, and the results should be interpreted as a snapshot rather than a stable property of model classes.

\subsection{Implications and future directions}

These findings have implications for both methodological evaluation and model development. The presence of structured interference in some models suggests that controlled experimental paradigms from psychology can be meaningfully adapted to probe LLM behavior. At the same time, the attenuation or absence of interference in other models indicates that such effects are not inevitable and may be mitigated through training or alignment.

Future work should extend this framework to additional domains, examine the role of prompt design and decoding strategies, and investigate the mechanisms underlying model-dependent differences. In particular, linking interference patterns to specific training or alignment procedures would provide deeper insight into how associative structure emerges in LLMs.

In summary, we find that associative interference under IAT-style constraints is present in some large language models but not others. By separating compliance from interference and modeling item-level variability, we present a controlled experimental framework for isolating associative interference under constrained response conditions. These results highlight the importance of model-specific evaluation and caution against treating associative behavior as a uniform property of contemporary language models.

\section{Conclusion}

Evaluating bias in large language models requires methods that go beyond surface-level inspection of outputs. By adapting a validated psychological instrument and applying trial-level inferential analysis, this study demonstrates that large language models differ in whether they express structured associative asymmetries under controlled conditions. Both positive and null findings are informative when interpreted within a principled measurement framework.

These results highlight the value of psychology-informed approaches for understanding how social regularities are encoded and expressed in artificial systems, and caution against equating uniform or neutral-seeming outputs with the absence of bias. Measuring how associations manifest is as important as determining whether they appear at all.



\section{Data availability}

Data are archived in Zenodo and will be made publicly available upon publication [dataset] \cite{cohen2026dataset}.
A restricted link is here for reviewer access:
\href{https://zenodo.org/records/19557680?preview=1&token=eyJhbGciOiJIUzUxMiJ9.eyJpZCI6ImI3MzlkOTNlLTg2YjMtNDVmNi05NDBmLTYzNTE3NzFjMzJhZiIsImRhdGEiOnt9LCJyYW5kb20iOiIyYjFkZTZmZjRmMDFjN2NlMTNjNjI1YmJiZTI3ZTA3NiJ9.DUdxHNTorOCPtCxsDjmOavof-ZLgrHAh0cGBpOmkFr0xj3TPtp5vf6sHd9BdxatSvHM74AYXnXLu34CX73ThEAhere}{HERE}.









\bibliography{main}
\bibliographystyle{tmlr}

\appendix
\section{Appendix}

\subsection*{S1. Compliance Model Specification}

Compliance with the forced-choice response format was modeled using hierarchical Bayesian logistic regression. The binary outcome variable indicated whether a model produced a valid structured response (``A'' or ``B'').

Let $Y_{ijkl}$ denote compliance for trial $i$ from model $j$, IAT domain $k$, block $l$, and item $m$. The model was specified as:

\begin{equation}
    Y_{ijkl} \sim \text{Bernoulli}(p_{ijkl})
\end{equation}
\begin{align}
     \text{logit}(p_{ijkl}) &=
\beta_0
+ \beta_1 \text{Block}_{l}
+ \beta_2 \text{Model}_{j} 
+ \beta_3 \text{Domain}_{k}+ \\ \nonumber
&\beta_4 (\text{Block} \times \text{Model})
+ \beta_5 (\text{Block} \times \text{Domain})
+ \\ \nonumber
&\beta_6 (\text{Model} \times \text{Domain})
+ \\  \nonumber
& \beta_7 (\text{Block} \times \text{Model} \times \text{Domain})
+ u_m,
\end{align}

where $u_m \sim \mathcal{N}(0, \sigma_{\text{item}})$ represents item-level random intercepts capturing stimulus-specific compliance heterogeneity.

\subsection*{S2. Prior Specification}

Weakly regularizing priors were used to stabilize estimation while allowing the data to dominate inference.

Fixed effects were assigned:

\[
\beta_r \sim \mathcal{N}(0, 1)
\]

reflecting the expectation that large log-odds effects are unlikely but not impossible.

The intercept was assigned:

\begin{equation}
    \beta_0 \sim \mathcal{N}(3, 1.5)
\end{equation}

corresponding to a broad prior expectation of high compliance (centered near 95\% probability) without enforcing near-deterministic behavior.

The item-level standard deviation was assigned:

\begin{equation}
\sigma_{\text{item}} \sim \text{Exponential}(1)
\end{equation}

a weakly informative prior favoring moderate variability while permitting larger dispersion if supported by the data.

\subsection*{S3. Posterior Estimation}

Models were fit using Hamiltonian Monte Carlo (No-U-Turn Sampler) as implemented in \texttt{brms} and Stan. Four chains were run for 4,000 iterations each, with 1,500 warm-up iterations per chain, yielding 10,000 post-warmup draws.

Posterior summaries are reported as means with 95\% credible intervals, in Table \ref{tab:compliance_compact}. Raw observed rates are provided in Table~\ref{tab:compliance_observed}.

\begin{table}[ht]
\centering
\caption{Posterior predicted compliance probability (\%) by model and IAT domain. Values are means with 95\% credible intervals.}
\label{tab:compliance_compact}
\begin{tabular}{llll}
\toprule
Model & IAT Domain & Congruent (\%) & Incongruent (\%) \\
\midrule
Claude Sonnet-4 & Gender-Career  & 99.8 [99.3, 100.0] & 99.9 [99.4, 100.0] \\
Claude Sonnet-4 & Gender-Science & 99.7 [99.1, 100.0] & 99.8 [99.2, 100.0] \\
\addlinespace
Gemini 2.5 Pro  & Gender-Career  & 99.9 [99.5, 100.0] & 99.9 [99.5, 100.0] \\
Gemini 2.5 Pro  & Gender-Science & 99.9 [99.4, 100.0] & 100.0 [99.6, 100.0] \\
\addlinespace
GPT-5           & Gender-Career  & 98.4 [96.9, 99.4]  & 98.6 [97.1, 99.5] \\
GPT-5           & Gender-Science & 98.1 [96.5, 99.2]  & 98.3 [96.8, 99.3] \\
\bottomrule
\end{tabular}
\begin{tablenotes}
\small
\item Compliance is defined as the production of a valid forced-choice response (A or B). Estimates derived from hierarchical Bayesian logistic regression with item-level random intercepts.
\end{tablenotes}
\end{table}


\subsection*{S4. Convergence Diagnostics}

Convergence diagnostics indicated satisfactory model performance. All parameters exhibited $\hat{R} \leq 1.01$, indicating adequate mixing across chains. Effective sample size ratios exceeded conventional thresholds for stable estimation. No divergent transitions were observed in any chain.

Posterior predictive checks indicated appropriate model fit.

\begin{table}[ht]
\centering
\begin{threeparttable}
\caption{Observed compliance rates (valid A/B responses) by model and condition.}
\label{tab:compliance_observed}
\begin{tabular}{lllll}
\toprule
Model & IAT Domain & Block & Valid / Total & Observed (\%) \\
\midrule
Claude Sonnet-4 & Gender-Career  & Congruent   & 80 / 80 & 100.0 \\
Claude Sonnet-4 & Gender-Career  & Incongruent & 79 / 80 & 98.8 \\
Claude Sonnet-4 & Gender-Science & Congruent   & 80 / 80 & 100.0 \\
Claude Sonnet-4 & Gender-Science & Incongruent & 79 / 80 & 98.8 \\
\addlinespace
Gemini 2.5 Pro  & Gender-Career  & Congruent   & 80 / 80 & 100.0 \\
Gemini 2.5 Pro  & Gender-Career  & Incongruent & 80 / 80 & 100.0 \\
Gemini 2.5 Pro  & Gender-Science & Congruent   & 80 / 80 & 100.0 \\
Gemini 2.5 Pro  & Gender-Science & Incongruent & 80 / 80 & 100.0 \\
\addlinespace
GPT-5           & Gender-Career  & Congruent   & 75 / 80 & 93.8 \\
GPT-5           & Gender-Career  & Incongruent & 75 / 80 & 93.8 \\
GPT-5           & Gender-Science & Congruent   & 75 / 80 & 93.8 \\
GPT-5           & Gender-Science & Incongruent & 75 / 80 & 93.8 \\
\bottomrule
\end{tabular}
\end{threeparttable}
\end{table}


\subsection{Conditional Interference Model}

For valid forced-choice responses, task consistency was defined as:

\begin{equation}
   Y_{ijkl} =
\begin{cases}
1, & \text{\small if LLM selection matched 
the stimulus's predefined domain}\\
0, & \text{otherwise}
\end{cases} 
\end{equation}

and modeled as:

\begin{equation}
    Y_{ijkl} \mid \text{valid}_{ijkl}=1 \sim \text{Bernoulli}(p_{ijkl})
\end{equation}

\begin{align}
 \text{logit}(p_{ijkl}) &= 
\beta_0
+ \beta_1 \text{Block}_{l}
+ \beta_2 \text{Model}_{j}
+ \beta_3 \text{Domain}_{k}
+ \\\nonumber & 
\beta_4 (\text{Block} \times \text{Model})
+ \beta_5 (\text{Block} \times \text{Domain})
+ \\\nonumber & \beta_6 (\text{Model} \times \text{Domain})
+ \\\nonumber & \beta_7 (\text{Block} \times \text{Model} \times \text{Domain})
+ u_m   
\end{align}

where, 

\[
u_m \sim \mathcal{N}(0,\sigma_{\text{item}})
\]

where $u_m$ denotes the item-level random intercept. Interference magnitude was summarized on the probability scale as
\begin{equation}
    \Delta P =
P(\text{consistent} \mid \text{congruent})
-
P(\text{consistent} \mid \text{incongruent}),
\end{equation}

such that positive values indicate lower task consistency in incongruent blocks.

\end{document}